%% file: EH.tex
\input ictp.tex

\hfuzz=50pt

\TITLE{EINSTEIN AND HILBERT:}
\TITLE{THE CREATION OF GENERAL RELATIVITY \FOOTNOTE{$^*$}
{Expanded version of a Colloquium lecture held at the International Centre for
Theoretical Physics, Trieste, 9 December 1992 and (updated) at the
International University Bremen, 15 March 2005.}}
\vskip1.5truecm

\AUTHOR{Ivan T. Todorov}
\centerline{Institut f\"ur Theoretische Physik, Universit\"at
  G\"ottingen, Friedrich-Hund-Platz 1}  
\centerline{D-37077 G\"ottingen, Germany; 
e-mail: itodorov@theorie.physik.uni-goe.de}
\centerline{and}
\centerline{Institute for Nuclear Research and Nuclear Energy,
  Bulgarian Academy of Sciences}
\centerline{Tsarigradsko Chaussee 72, BG-1784 
Sofia, Bulgaria;\FOOTNOTE{$^{**}$}{Permanent address.}  e-mail: 
todorov@inrne.bas.bg}
\vskip1truecm
\ABSTRACT

It took eight years after Einstein announced the basic physical ideas
behind the relativistic gravity theory before the proper mathematical
formulation of general relativity was mastered. The efforts of the greatest
physicist and of the greatest mathematician of the time were involved and
reached a breathtaking concentration during the last month of the work.

Recent controversy, raised by a much publicized 1997 reading of
Hilbert's proof-sheets of his article of November 1915, is also discussed.

\vfill\eject

\bigskip

{\bf Introduction}

Since the supergravity fashion and especially since the birth of superstrings
a new science emerged which may be called ``high energy mathematical
physics''. One fad changes the other each going further away from accessible
experiments and into mathematical models, ending up, at best, with the
solution of an interesting problem in pure mathematics. The 
realization of the grand original design seems to be, decades later, 
nowhere in sight. For quite some time, though, 
the temptation for mathematical physicists (including leading 
mathematicians) was hard to resist. Yuri Manin characterized the
situation as ``an extreme romanticism of the theoretical high energy 
physics of the last quarter of our century''.

There does exist, on the other hand, a true example of a happy competition
between mathematics and physics which has led to the most accomplished 
among the three claimed revolutions in our science in the first
quarter of the twentieth century: the creation of the general 
theory of relativity. It illustrates how difficult it has been 
- even for the founding fathers of the theory - to fully understand
and adopt such basic notions as reparametrization invariance, Bianchi 
identities, the concept of energy, which
nowadays enter a student curriculum. The presence of a controversy - if
not so much among the participants in the events, at least among the 
historians of science nearly a century later - could serve one good 
purpose: to attract a wider audience to this remarkable story.  

\bigskip

\noindent{\bf 1.\ \ Prologue: Einstein (and Grossmann): 1907--1915}

Einstein seemed never happy with what he had achieved. He was not
satisfied by the special principle of relativity because it did not
incorporate accelerated motion. Since his student years he had
absorbed with sympathy Ernst Mach criticism of the ``monstrous
[Newtonian] notion of absolute space''. (It was his lifelong friend
Michele Besso who induced the 18 year old Einstein -- back in 1897 -- to
read Mach's ``History of Mechanics''.)

In his recently reprinted address to the 1904 International Congress
of Arts and Science St. Louis, \FOOTNOTE{$^{*)}$}{H. Poincar\'e, L'
  \'etat actuel et l'avenir de la physique math\'ematique, {\it
    Bulletin des Sciences Math\'ematiques} {\bf 28} (1904) 302-324;
  reprinted as: The
  present and future of mathematical physics, {\it Bull. Amer. Math. 
Soc.} {\bf 37} (2000) 25-38.}\ as well as in his fundamental paper 
``Sur la dynamique de l'\'electron'' which appeared in 1906, Poincar\'e 
already states the problem of modifying Newton's gravity theory in
order to make it consistent with relativity. The problems of gravity 
and of relativity of accelerated motion are combined in what Einstein 
will call 13 years later ``{\it the happiest thought in my life}'' 
(Pais 82, Chap.9, pp.178--179). Then, in 1907, while working on the 
review article ``\"Uber das Relativit\"atsprinzip und die aus
demselben gezogenen Folgerungen'' (``On the principle of relativity and 
its consequences''), Jahrbuch der Radioaktivit\"at und Elektronik {\bf
  4} (1907) 411--461 (written, ironically, on the request of Johannes
Stark who appears as a bitter enemy of relativity during the Nazi
period), he has the idea that 
``{\it for an observer falling freely from the roof of a house there
exists} -- at least in his immediate surroundings -- {\it no 
gravitational field}''. In his Kyoto lecture (cited by Pais) Einstein
recalls: ``I was sitting in a chair in the patent office at Bern when 
all of a sudden a thought occurred to me: `If a person falls freely he 
will not feel his own weight'. I was startled . . . ''

Thus the celebrated equivalence principle first appears just two
years after the formulation of special relativity (although it was
only so baptized by Einstein 5 years later). The 1907 paper does not
stop at that. It contains a derivation of the gravitational red
shift. Einstein also deduces the formula 
$c(\phi )=c\left( 1+{\phi\over c^2}\right)$ 
for the velocity of light along the direction $\xi$ of a constant 
gravitational field (the mass in the gravitational potential $\phi$ 
being identified with the unit of mass 
so that $\phi /c^2$ appears to be dimensionless). He infers 
that \FOOTNOTE{$^{*)}$} {In predicting the bending of light Einstein 
has a prominent predecessor. In the first ``Query'' to his 
``Opticks'' Isaac Newton writes: ``Do not Bodies act upon Light at a 
distance, and by their action bend its  Rays; and is not this action 
strongest at the least distance?"}\ ``the light rays which do not run 
in the $\xi$ direction are bent by the gravitational field''. As if
all this was not enough for a first probe into relativistic gravity, Einstein
wrote to his friend Konrad Habicht on Christmas 1907 (just 3 weeks after
submitting the paper): ``I hope to clear up the so--far unexplained secular
change of the perihelion length of Mercury . . .'' (Pais 82, p.182).

All three observational implications of general relativity were in the
mental view of its creator already at this preliminary stage. Einstein's
genius is here manifest with all its flare as well as with its limitations.
The limitations are most honestly described by Einstein himself. In his
``Autobiographische Skizze'' completed in March 1955, a month before his
death, he deplors his attitudes towards advanced mathematics during his
student years. (Maurice Solovin, Einstein's close friend during the period
just after he graduated from the ETH -- the Z\"urich Polytechnic,
remembers that ``Einstein . . . often spoke against abusive use of
mathematics in physics. Physics, he would say, is essentially a concrete
and intuitive science.'' ``I do not believe in mathematics'', Einstein is
reported to have affirmed before 1910 -- see Pyenson 85, p.21, and
references cited there.) Already in 1907 Einstein is striving for a
generally covariant theory, but he is not aware that such a theory, the
Riemannian geometry, has been created in the 19th century. The (local)
equivalence principle purporting to generalize ``the happiest thought'' of
1907 to inhomogeneous fields is instrumental in Einstein's tracing the
road to identifying gravity with space time geometry. Yet, nowadays, a
mathematically minded student of relativity feels embarrassed if he has to
explain what such an ``equivalence principle'' does actually mean. The
presence of a non--zero gravitational field strength is manifested by a
non--zero curvature tensor and cannot be eliminated -- not even locally --
by a coordinate transformation -- no matter what acceleration one chooses.
(In order to make our point in a few words we are oversimplifying matters.
A detailed treatment of Einstein's principle of equivalence is contained in
(Norton 86).)

To summarize: by Christmas 1907 Einstein had all physical consequences of
the future gravity theory in his hands, yet, he had another eight
years to go and to appeal for mathematicians' help before arriving at 
the proper mathematical formulation of general relativity.

At first, though, Einstein behaved much like Michael Atiyah's physicist who
``not being able to solve a problem moves on to the next more difficult one''.
For three and a half years, from 1908 to mid 1911, Einstein's main
preoccupation was quantum theory: light quanta, blackbody radiation. (In
1910 he also completed a paper on critical opalescence, his last major
work on classical statistical physics.) Characteristically this was before
the early (1913) work of Niels Bohr after which quantum theory started
becoming popular. (Even later, in 1915, Robert Millikan, who spent 10 years
to test Einstein's prediction of the photoeffect will write: ``Einstein's
photoelectric equation . . . appears in every case to predict exactly the
observed results . . . Yet the semicorpuscular theory by which Einstein
arrived at his equation seems at present wholly untenable.'' -- see Pais 82,
p.357.) Einstein's own appraisal of his 1909--1910 assault on the light
quantum problem was not complimentary either. (In today's perspective, though,
he appears as a true pioneer in the quantum theory of those days -
see, e.g., Sec.2 {\it Einstein and the early quantum theory} in M.J. Klein's
lectures ``The Beginning of Quantum Theory'' in {\it History of
  Twentieth Century Physics}, Proceedings of the International School
of Physics ``Enrico Fermi'', Course LVII, ed. by C. Weiner, Academic
Press, New York and London 1977, pp.19-39.)

By June 1911, after a four month stay in Prague, Einstein is again on the
general relativity track and on the bending of light. The next important
breakthrough comes a year later, in August 1912 when Einstein is back in
Z\"urich and is literally crying for help to his friend and fellow student
from ETH: ``Grossmann, you must help me or else I'll go crazy!'' (Since 1907
Marcel Grossmann was a professor of geometry in ETH.) As witnessed in
Einstein's correspondence (letter to L. Hopf of 16 August, 1912) at this time
the two of them understood that gravity should be described not by a single
scalar field (which Einstein related in previous publications with the
variation of the velocity of light), but by the symmetric tensor metric
field $g_{\mu\nu}(x)$ which has 10 independent components. Grossmann
quickly realized that the generally covariant formalism Einstein was
looking for (equivalence of arbitrarily moving frames) was provided by
Riemannian geometry. Yet, this was only the beginning of the hard work.
Einstein was absorbed to a point at which he refused to talk about quantum
theory anymore. On 29 October 1912 he wrote to Sommerfeld: ``I assure you
that with respect to quantum I have nothing new to say . . . I am now
exclusively occupied with the problem of gravitation and I hope to master
all difficulties with the help of a friendly mathematician here. But one
thing is certain: in all my life I have labored not nearly as hard, and I have
become imbued with great respect for mathematics, the subtler part of
which I had in my simple--mindedness regarded as pure luxury until now.
Compared with this problem the original relativity is a child's play.'' (Mehra
73, p.93 and Pais 82, p.216).

It appears that in the subsequent months Einstein still trusts better his
(uncommon!) physical intuition, rather than the mathematical wisdom. In
their first joint paper with the glorious title: ``Entwurf einer
verallgemeinerten Relativit\"atstheorie und einer Theorie der Gravitation. I.
Physikalischer Teil von Albert Einstein. II. Mathematischer Teil von Marcel
Grossmann'' (Leipzig und Berlin, B.G. Teubner 1913; reprinted with added
``Bemerkungen'' in Zeitschrift f\"ur Mathematik und Physik {\bf 62} (1914)
225--261). Grossmann notes in the mathematical part that the Ricci tensor
$R_{\mu\nu}$ may be used for the formulation of a generally covariant
gravity theory - an important step towards the ultimate formulation of
the basic equation of general relativity. (As stressed in (Win 04) 
Grossmann deserves more credit for this achievement than usually 
given.) But the authors reject this possibility, since it allegedly 
violates ``the physical requirements''. The crucial mistake comes from 
Einstein's ``causality requirement'': the metric tensor $g_{\mu\nu}$ 
should be completely determined from the stress energy tensor. This is 
certainly not correct for the true equations of general relativity,
$$G_{\mu\nu}=\kappa\ T_{\mu\nu}\quad {\rm where}\quad
G_{\mu\nu}:=R_{\mu\nu}-{1\over 2}\ R\ g_{\mu\nu},  \eqno(1)$$
first reported by Hilbert in his paper submitted (to
Nach. Ges. Wiss. G\"ottingen), 20 November 1915.
Indeed, $G_{\mu\nu}$ satisfies the Bianchi identities
$(G^{\mu\nu})_{;\nu}=0$ in accord with the (covariant) energy momentum
conservation law. Hence, only six of its ten components are independent, so
that the $g_{\mu\nu}$, far from being uniquely determined, depend on four
arbitrary functions. We now, sure, understand what it means. General
covariance says that the choice of coordinates is a matter of convention
which should not affect physics. The metric tensor much like the
electromagnetic potential is not an observable. To determine it one needs
(on top of $T_{\mu\nu}$) four ``coordinate conditions'' corresponding to the
gauge fixing in electrodynamics. It would be too easy to criticize Einstein
on the ground of knowledge acquired by physics decades later. (The notion of
gauge invariance (Eichinvarianz) first appears six years later in
Hermann Weyl's ``Gravitation und Elektrizit\"at''
(Sitzungsber. d. Preuss. Akad. d. Wiss. (1918) 465-478)\FOOTNOTE
{$^*)$} {After over half a century Paul Dirac (Proc. Roy. Soc. {\bf
  A333} (1973) 403-418) still views this paper of Weyl as ``unrivalled
by its simplicity and beauty''.}\ describing a hypothetical dilation
symmetry - in an early attempt to construct a unified field theory. It made
its way to where it really belongs -- Maxwell--Dirac electrodynamics --
again thanks to Weyl after another 10 years. In 1912--1915 Einstein was
well ahead of his time exploring, in the words of Pais, a ``no man's land''.)

In short, in the ``Entwurf'' (``Outline'') Einstein and Grossmann back
down  from general covariance and settle for a set of not quite
geometric equations only invariant under linear coordinate
transformations. Einstein is not happy with it. In August 1913 he
writes to Lorentz: ``The gravitational equations unfortunately do not
have the property of general covariance . . . However, the whole faith 
in the theory rests on the conviction that acceleration of the
reference system is equivalent to a gravitational field. Thus, if not all
equations of the theory . . . admit transformations other than linear ones,
then the theory contradicts its own starting point . . . all is then
up in the air'' (Pais 82, p.228).

In early 1914 Einstein and Adriaan Focker (who had just received his Ph.D
under Lorentz) restored general covariance but at a high price. They derived
the scalar equation $R=-\kappa\ T(R=R^\nu\ _\nu ,T=T^\nu\ _\nu )$
assuming that the metric is conformally flat,
$g_{\mu\nu}=\psi^2\eta_{\mu\nu}$, i.e., returning, essentially, to the
scalar theory of gravity. In October 1914 Einstein completes a 56-page long
paper ``Die formale Grundlage der allgemeinen Relativit\"atstheorie'' which
goes back to the Einstein--Grossmann theory. (Einstein's strength is
not in the mathematical formalism: his 1914 treatment of the covariance
properties of the field equations, criticized by Levi-Civita, would
hardly have impressed educated geometers - see Sect. 4 and footnote 124 of
(Sau 99); it does not please its author either.) In the beginning
of 1915 Einstein appears fed up (if not fully satisfied) with general
relativity, and he goes ahead to do some experimental work with the Dutch
physicist Wander de Haas (they discover a new effect: the torque of a
suspended iron cylinder as a consequence of an abrupt magnetization).
\bigskip

\noindent{\bf 2.\ \ Berlin -- G\"ottingen (1915)}\FOOTNOTE {$^*)$} {A
well researched and lively account of Einstein's Berlin years is
provided in (Goen 05); (Reid 96) is a standard source for Hilbert's
life (1862-1943), most of which (since 1895) is spent in G\"ottingen.}\
 
On 29 November 1971 Eugene Wigner writes to Jagdish Mehra asking him one
of those questions, people, who have come to know Wigner, can easily
imagine: ``. . . I was under the impression that, simultaneously with
Einstein, Hilbert also found the now accepted equations of general
relativity. Is this correct? If so, is there a reason no one seems to mention
this now? I realize that the basic idea was due to Einstein but it is
interesting that, even after the promulgation of the basic idea, it took a
rather long time to find the correct equations incorporating that idea --
even though both Einstein and Hilbert seem to have worked on it.'' Mehra
replies to Wigner within two weeks by a long letter and later publishes 
an 87-page paper on the subject (Mehra 73). But the real answer to
Wigner's question comes another five years later in an article by Earman and
Glymour who have digged into the Einstein Papers at Princeton University.

David Hilbert, whose 23 ``Honors Class'' problems (Yan 02) occupy
mathematicians throughout the 20th century, is, in his fifties (after
Poincar\'e, 58, dies in 1912), the uncontested leader of the
world of mathematics. Having published a (by now, classic) book, ``{\it
Foundations of Geometry}'' Hilbert states his sixth problem (Paris,
1900): ``To treat in the same manner, by means of axioms, those
physical sciences in which already today mathematics plays an
important part...''. His lifelong belief that every scientific
problem can - and will - be solved is reflected in the words engraved
on his tombstone in G\"ottingen: ``Wir mussen wissen, wir werden
wissen.'' (``We must know, we shall know.''). Starting with 1912, 
after completing his book 
on linear integral equations, Hilbert's main preocupation becomes
mathematical physics: the realization of the program encoded in his 
sixth problem. He thinks of unifying within the axiomatic approach the 
new electromagnetic theory of the electron, put forward in 1912 by 
Gustav Mie (1869-1957), with the Einstein-Grossmann theory 
(reported to the G\"ottingen Mathematical
Society in 1913, shortly after its publication - see (Sau 99),
Sect. 2.1). Hilbert tries to have Einstein visiting G\"ottingen 
(he invites him more than once - first in 1912) and this time            
he succeeds.

In late June - early July 1915 Einstein spends a week in G\"ottingen where (as
he witnesses in a letter to Zangger of 7 July) he ``gave six two--hour lectures
there''. By all accounts he seems happy with the outcome: ``To my great
joy, I succeeded in convincing Hilbert and Klein completely'' (E. to de Haas)
``I am enthusiastic about Hilbert'' (E. to Sommerfeld). The feelings appear
to be mutual. Hilbert recommends Einstein for the third Bolyai Prize in
1915 for ``the high mathematical spirit of his achievements'' (the first and
the second recipients of the Bolyai prize have been Poincar\'e and Hilbert
-- see Mehra 73).
Nevertheless, the G\"ottingen discussions seem to have reinforced Einstein's
uneasiness about the lack of general covariance of his (and Grossmann's)
equations. He is reluctant (he writes to Sommerfeld in July 1915) to
include his papers on general relativity in a new edition of ``{\it The
Principle of Relativity}'', ``because none of the presentations to date is
complete''. After the November race Einstein will state more
precisely (in letters to friends) the grounds for his discontent with the old
theory: (1) its restricted covariance did not include uniform rotations; (2)
the precession of the perihelion of Mercury came out 18$''$ instead of the
observed 45$''$ per century; (3) his proof of October 1914 of the uniqueness
of the gravitational Hamiltonian is not correct.

In the meantime Einstein receives a letter by Sommerfeld (perhaps in late
October 1915 -- the letter is lost) from which he learns that he is not the
only one dissatisfied with his 1914 theory. Hilbert also has objections to it
and is working on his own on ``Die Grundlagen der Physik'' originally
conceived as ``Die Grundgleichungen /basic equations/ der Physik'' - see
(Sau 99) footnotes 73 and 90). Will Einstein let someone else, be it 
Hilbert himself, share with him the fruit of years of hard work
and great inspiration? Not he! At 36, he can still fight. The Einstein
papers reveal an unprecedented activity in November 1915.

Einstein submits four communications to the ``Preussische Akademie der
Wissenschaften'': on 4, 11, 18 and 25 November -- no Thursday is skipped!
These are not different parts of a larger work. The first, ``Zur allgemeine
Relativit\"atstheorie'' rejects his formulation of 1914 and proposes a new
fundamental equation. The second, with the same title, rejects the first
and starts anew. The fourth, ``Die Feldgleichungen der Gravitation'' rejects
the first two and finally contains the right equations. It is like in a movie
when the film is turned on a high speed. Nothing similar has happened
either before or after in Einstein's life.

But this is not all. Einstein only answers (the lost) Sommerfeld's
letter on 28 November (three days after his last talk at the Academy). 
``Don't be angry with me'' he writes ``for only today answering your 
friendly and interesting letter. But last month I had one of the most 
exciting, most strenuous times of my life,
also one of the most rewarding. I could not concentrate on writing''. Indeed,
from late October to late November Einstein stops writing to any of his
habitual addressees: Besso, Ehrenfest, Lorentz, . . . But he does write
letters (or, rather, postcards). He only replaces all his regular
correspondents by a single new one -- Hilbert. Four postcards are preserved
from Einstein to Hilbert dated 7, 12, 15, 18  November and two of the four
Hilbert answers.

On 7 November Einstein sends to Hilbert the proofs of his
November-four paper and in the accompanying card writes 
``I recognized four weeks ago that my
earlier methods of proof were deceptive''. He alludes to the above mentioned
letter of Sommerfeld which reports on Hilbert's objections to the October
1914 paper; and closes by saying: ``I am curious whether you will be well
disposed towards this solution''.

Hilbert would have hardly been well disposed towards the new equation,
since it assumes that the determinant of the metric tensor is a constant
(-1) and is hence still not generally covariant. Probably, after having
Hilbert's criticism (which has been lost) Einstein opted on 11 November for
the generally covariant equation
$$R_{\mu\nu}=\kappa\ T_{\mu\nu}       \eqno(2)$$
which Grossmann and he have rejected two years earlier. It only coincides,
however, with the correct equation (1) if $T_{\mu\nu}$ (and hence also
$R_{\mu\nu}$) is traceless. This is the case of Maxwell electrodynamics
and Einstein speculates that it may be more general.

The next day, 12 November, Einstein sends a second postcard to Hilbert
announcing that he had finally achieved generally covariant field equations.
He also thanks Hilbert for his ``kind letter'' (which is lost). Hilbert replies
on 14 November a long message on two postcards. He is excited about his
own ``axiomatic solution of your grand problem''. In a postscript Hilbert adds
that his theory is ``wholly distinct'' from Einstein's and invites Einstein to
come to G\"ottingen and hear his lecture on the subject. The tone is cordial:
Hilbert urges Einstein to come to G\"ottingen the day before the lecture and
pass the night at Hilbert's home. The next day, Monday, 15 November,
Einstein already answers Hilbert's cards. (One cannot fail to notice how
accurately the mail service is working in Germany in the midst of the
European war.) ``The indications on your postcards lead to the greatest
expectations''. He apologizes for his inability to attend the lecture, 
since he is overtired and bothered by stomach pains. 
Asks for a copy of the proofs of
Hilbert's paper. Apparently, he does receive the requested copy within 
three days, because on 18 November, the day of his third talk at the
Academy, Einstein writes his fourth postcard: ``The system [of equations]
given by you agrees -- as far as I can see -- exactly with what I found in
recent weeks and submitted to the Academy''. Then Einstein remarks that he
has known about Eq.(2) ``for three years'' but that he and Grossmann have
rejected it on the grounds that in the Newtonian limit they are not
compatible with ``Newton's law'' (meaning Poisson's field equation). Finally,
Einstein informs Hilbert that he is finally explaining the advance of
the perihelion of Mercury from general relativity alone without the aid of
any subsidiary hypotheses.

Two remarks are in order.

First, it is not true that Hilbert's Eq.(1) is equivalent to Einstein's
Eq.(2) of the paper submitted to the Academy on 11 November. (It will be
equivalent to the equation Einstein is going to write a week later. It seems,
however, that Einstein does have in mind his Eq.(2) in this postcard since he
is adding the priority claim that he knew it for three years.) The two 
equations are only consistent with one another for $T(=T^\nu_\nu )=0$,
the case Einstein has been mostly interested in at the time.

Second, Einstein does derive the correct value for the advance of the
perihelion of Mercury in his third communication ``Erkl\"arung der
Perihelbewegung des Merkur aus der allgemeinen Relativit\"atstheorie''
from his not exactly correct equation. This is possible since he is
actually solving the homogeneous equation (with $T_{\mu\nu}=0$) in the
post Newtonian approximation (allowing for point singularities). - In
seeing the physical implications of the theory Einstein has no competitor.

The next day, Friday the 19th, Hilbert congratulates Einstein for having
mastered the perihelion problem and adds cheerfully: ``If I could calculate
as quickly as you, then the electron would have to capitulate in the face of
my equations and at the same time the hydrogen atom would have to offer
its excuses for the fact that it does not radiate'' (Pais 82, p.260).

On 20 November Hilbert presents to the Gesellschaft der Wissenschaften in
G\"ottingen his work. He {\it derives} the correct equations from the
variational principle assuming general covariance (we would say today
reparametrization invariance) and a second order equation for
$g_{\mu\nu}$. He gives full credit to Einstein's ideas. On the first page of
his article he writes: ``Einstein . . . has brought forth profound thoughts and
unique conceptions, and has invented ingeneous methods for dealing with
them . . . Following the axiomatic method, in fact from two simple axioms, I
would like to propose a new system of the basic equations of physics. They
are of ideal beauty and I believe they solve the problems of Einstein and Mie
at the same time''. In the published version Hilbert refers to all Einstein
November papers. About the one of 25 November, submitted after his talk,
he says: ``It seems to me that [our] differential equations of gravitation are
in agreement with the noble theory of general relativity proposed by
Einstein in his later memoire''.

On 25 November Einstein proposes {\it without derivation} the equation
$$R_{\mu\nu}=\kappa \left( T_{\mu\nu}-{1\over 2}\ T\ g_{\mu\nu}
\right)            \eqno(3)$$
which is exactly equivalent to Hilbert's Eq.(1), since they both imply
$R+\kappa\ T=0$. He chooses not to mention Hilbert's name in the published
paper. Later commentators have a hard time to understand what was
Einstein's argument at the time to include the trace term. Only Norton
makes a well documented (59 pages long) case 
(including the study of a Z\"urich notebook
of Einstein) for an independent Einstein's road to the correct equations.
\bigskip

\noindent{\bf Aside:}\quad
Today's student will easily find the $-{1\over 2}\ R$ term (or equivalently,
the $-{1\over 2}\ T$ term) using the Bianchi identity. The trouble was,
Einstein did not know them. We should not be too hard on him on that
account. Hilbert, too, does not know them: he derives four identities 
among the fields in his theory - anticipating, three years in advance,
a special case of Noether's theorem (Viz 94) (complemented with a
stronger statement - see (Sau 99), Sect. 3.3 and footnote 120) but 
he conjectures
erroneously that they will enable him to express the electromagnetic
potential in terms of the gravitational field. He corrects his error in a
later version of the paper.

Felix Klein who (in 1918) reduces the vanishing of the covariant 
divergence,

$$\left( R^{\mu\nu}-{1\over 2}\ g^{\mu\nu}\ R\right)_{;\nu}=0,$$
to Noether's theorem as well, does not realize that it is a
consequence of
the Bianchi identities for the Riemann curvature tensor either.  
``Bianchi identities'' are known before Bianchi to
Aurel Voss (1880) and to Ricci (1889); Luigi Bianchi (a pupil of Klein's!)
rediscovers them in 1902.
\bigskip

\noindent{\bf 3.\ \ Aftermath. Controversy among historians of science}

In his speech on the occasion of Planck's 60$^{th}$ birthdate (in 1918)
Einstein talks about different categories of people that have devoted
themselves to science. For some science is a sport which allows them to
satisfy their pride or vanity. If the angel chases all such people from the
temple of science, he continues, then the temple would remain almost
empty, but Planck will be among the  precious few who will remain.

The chronicle of the last month of the creation of general relativity
demonstrates that the spirit of competition has not been alien to Einstein
himself (as it was not to Leibniz and Newton). It is to the credit of both
Hilbert and Einstein that their November 1915 rivalry did not grow into a
public argument. Yet the outcome of the November events resulted in some
uneasy feelings between the two men. On 20 December 1915, Einstein writes
to Hilbert: ``I want to take the opportunity to say something to you which is
important to me. There has been a certain spell of coolness between
us, the cause of which do not want to analyze. I have, to be sure, 
struggled against any resentment, and with complete success. I 
think of you once again with untroubled friendliness, and ask you to 
try to think of me in the same way. It is really
a shame when two such real fellows [zwei wirkliche Kerle], whose work
has taken them above the shaby world, give one another no pleasure''. (EG
78, p.306; a slightly different reading of the German original the reader
will find in Pais 82, p.260.)

In his expository paper ``Die Grundlage der allgemeinen
Relativit\"atstheorie'', Annalen der Physik {\bf 49}, 769-822 (received 20
March, 1916) Einstein already refers (albeit superficially) to Hilbert's work.
In May 1916 he gives a colloquium in Berlin on Hilbert's paper. On that
occasion he again writes to Hilbert asking him to
explain his work (and complaining about its obscurity).

Hilbert's appreciation of Einstein appears unequivocal. His biographer, 
(Reid 96), attributes to him the words: ``Every boy in
the streets of G\"ottingen understands more about four--dimensional
geometry than Einstein. Yet, . . . Einstein did the work and not the
mathematicians''.

The story does not end here, however: it is continued by the next
generation of Einstein biographers and students of science history.

In 1997 a noteworthy addition to existing Einstein's biographies, (FL 97), 
appeared in English, providing a nice complement to (Pais 82).

Summing up the decisive phase of his work on general relativity
(Fl 97) quotes Einstein's letter to Heinrich Zangger (see also an 
earlier discussion of this letter in (Med 84)) which says: 
``{\it Only one colleague truely understood it, and he
 now tries skillfully to 'nostrify' it}'' [i.e. appropriate ('make it ours')]. 
We already know that the colleague in question was none other than
David Hilbert. F\"olsing justly refutes the accusation on the basis
of available evidence.

Later the same year an article in the 14 November issue of {\it
  Science}, (CRS 97) made the news. This paper has a direct bearing on 
our topic. It points out that a lately discovered proof-sheet of
Hilbert's paper, with a publisher's stamp of 6 December 1915, i.e. after the
publication of the fourth of Einstein's communications, involves substantial
changes in the manuscript. The fact that Hilbert modified his paper
after its submission has been known before: as we noted he had cited
all four Einstein's November papers and had commented on
the last one (submitted after his) in the published version of his
November 20 article. The authors strive to attribute a
great significance to the fact that the original text only involves
the Hilbert action, while the field equations, which are derived from
it, appear to be first inserted at the stage of the proofreading. Their
attempt to support on this ground Einstein's accusation of
``nostrification'' goes much too far. A calm, non-confrontational
reaction was soon provided by a thorough study (Sau 99) of Hilbert's 
route to the ``Foundations of Physics'' (see also the relatively even 
handed survey (Viz 01)). A direct critical comment on the unfounded
accusations in (CRS 97), (Win 04), originally rejected by the editors 
of {\it Science}, \FOOTNOTE {$^*)$}
{There seems to be a concerted effort to present the view of (CRS 97)
 as a final generally accepted ``Decision''. In a little more than a
 3-page long article on Hilbert (by J.J. O'Connor and E.F. Robertson),
available at
http://www-groups.dcs.st-and.ac.uk/~history/Mathematicians/Hilbert.html, 
the authors have found it necessary to devote a
paragraph citing the (CRS 97) accusation of inappropriate
behaviour against Hilbert, preceded by the words ``the authors /of
(CRS 97)/ show convincingly ...''. It is this type of overtly prejudiced
attitude that provokes uncommonly angry reactions as (LMP 04) and
gives credibility to extremist publications as (Bje 03).}\
finally appears in a more specialized journal (Win 04). 
\FOOTNOTE {$^**)$} {In their
  ``Response'' (CRS 04) to this comment the authors of (CRS 97)
  continue to assert that taking the variational 
derivative of the Hilbert action (a routine 3-line exercise for an 
average graduate student) is something Hilbert was not able to do by 
himself in 1915, and even compare it with the calculation of ``the 
one-billionth digit of {$\pi$}'' (that would require a supercomputer 
and the dedication of someone - like the Chudnovsky brothers (see
R. Preston, The Mountains of PI, {\it New Yorker}, March 2, 1992) - 
to program it).}\

The polemics is getting rough. A new book, (Wuensch 05), is advertised with 
a question mark: ``Ein Kriminalfall in der
Wissenschaftsgeschichte?'' (``A criminal case in the history of
science?''). The author asserts - already in the abstract to the book
- that a missing fragment (also discussed in (Sau 99) and in (Win 04)) 
of the text on 
pages 7 and 8 of Hilbert's proof-sheets, used in (CRS 97), contained 
``in all probability ... the explicit form of the field equations...'' 
She further argues that
``the passage ... was not excised originally but rather ... it must
have been deliberately removed in more recent times in order to
falsify the historical truth.''    

It is quite clear from  the November correspondence (and from recently
discovered letters of Max Born to Hilbert of the fall of 1915 - see
(Som 05)) - without appealing to criminal proceedings - that Hilbert's
competitive influence was crucial for Einstein's acceptance of general 
covariance - in spite of his long time reservations and doubts. The
analysis of the new evidence, detailed in (Sau 99) and in (Viz 01), 
indicates, on the other hand, that Hilbert appears to have been misled 
for a while, during the final race, in the opposite direction. 
After formulating the generally covariant action principle he appeals,
in his original text, to Einstein's long-promoted ``causality
principle'' and restricts the general covariance by a (non-covariant 
formulation of) the energy momentum conservation law. Only at the
stage of proofreading does Hilbert suppress all extra conditions and 
recognize the unqualified physical relevance of the covariant equation (1).

 Einstein and Hilbert had the moral strength and wisdom - after a 
month of intense competition, from which, in a final account,
everybody (including science itself) profited - to avoid a lifelong
priority dispute (something in which Leibniz and Newton failed). It
would be a shame to subsequent generations of scientists and
historians of science to try to undo their achievement. 
 
\ACK

An early version of this Colloquium lecture was held at the
International Centre for Theoretical Physics (Miramare-Trieste) in 
December 1992. Its text (which appeared as an ICTP internal report, 
IC/90/421) was prepared 
while the author was visiting at the Laboratorio Interdisciplinare 
per le Scienze Naturali ed Umanistiche of the International School 
for Advanced Studies (SISSA/ISAS) in Trieste. The present extended 
version of the March 15, 2005 Colloquium talk, held at the Physics 
Department of the International University Bremen, was written during
the author's visit as an Alexander von Humboldt awardee at the
Institut f\"ur Theoretische Physik, Universit\"at G\"ottingen. The
author is grateful to all these institutions for their hospitality 
and support. It is a pleasure to thank Hubert Goenner for providing
relevant recent references (including copies of unpublished Born's 
letters to Hilbert) and Karl-Henning Rehren for a critical reading of 
the manuscript. 

\bigskip

\vfill\eject

\centerline {REFERENCES}

\noindent (Bje 03) C.J. {\it Bjerkenes}, {\it Anticipations of Einstein
in the General Theory of Relativity} (XTX Inc. Downers Grove, Illinois
2003).

\noindent (CRS 97) Leo {\it Corry}, J\"urgen {\it Renn}, John {\it
  Stachel}, Belated decision in the Einstein-Hilbert priority dispute,
{\it Science} {\bf 278} (1997) 1270-1273.

\noindent (CRS 04) Leo {\it Corry}, J\"urgen {\it Renn}, John {\it
  Stachel}, Response to F. Winterberg, ``On 'Belated decision in the
Hilbert-Einstein priority dispute' published by L. Corry, J. Renn, and
J. Stachel'', anounced in a ``Retraction Notice'', {\it
  Z. Naturforsch.} {\bf 59a} (2004) 1004 at
http:www.mpiwg-berlin.mpg.de/texts/Winterberg-Antwort.html.
   
\noindent (EG 78) John {\it Earman}, Clark {\it Glymour}, Einstein and
Hilbert: Two months in the history of general relativity, {\it Arch. 
Hist. Exact Sci.} {\bf 19} (1978) 291--308.

\noindent (Fl 97) A. {\it F\"olsing}, {\it Albert Einstein: A
  Biography} (Viking, N.Y. 1997).

\noindent (Goen 05) Hubert {\it Goenner}, {\it Einstein in Berlin
  1914-1933} (C.H. Beck, M\"unchen 2005).

\noindent (LMP 04) A.A. {\it Logunov}, M.A. {\it Mestvirishvili},
V.A. {\it Petrov}, How were the Hilbert-Einstein equations discovered?
{\it Uspekhi Fizicheskikh Nauk} {\bf 174} (2004) 663-621 (English
translation: {\it Physics-Uspekhi} {\bf 47} (2004) 607-621); arXiv: 
physics/0405075.

\noindent (Med 84) H.A. {\it Medicus}, A comment on the relations
between Einstein and Hilbert, {\it Am. J. Phys.} {\bf 52}:3 (1984)
206-208.

\noindent (Mehra 73) Jagdish {\it Mehra}, Einstein, Hilbert and the theory of
gravitation in: {\it The Physicist's Conception of Nature}, edited by Jagdish
Mehra (D. Reidel Publ. Co., Dordrecht--Holland, Boston, USA 1973)
pp. 92--178.

\noindent (Norton 86) John {\it Norton}, What was Einstein's principle of
equivalence? pp. 5--47;
How Einstein found his field equations, 1912--1915, pp. 101--159, in {\it
Einstein and the History of General Relativity}, Don Howard, John Stachel,
eds., Einstein Studies, Vol.1 (Birkh\"auser, Boston, Basel, Berlin 1989).
Based on the Proceedings of the 1986 Osgood Hill Conference, North
Andover, Massachusetts.

\noindent (Pais 82) Abraham {\it Pais}, `{\it Subtle is the Lord . . .}', The
Science and the Life of Albert Einstein (Clarendon Press, Oxford 1982).

\noindent (Pyenson 85) Lewis {\it Pyenson}, {\it The Young Einstein}, The
advent of relativity (Adam Hilger Ltd., Bristol and Boston 1985).

\noindent (Reid 96) Constance {\it Reid}, {\it Hilbert} (Springer,
Berlin et al. 1996).

\noindent (Sau 99) Tilman {\it Sauer}, The relativity of discovery: 
Hilbert's first note on the Foundations of Physics, {\it Arch. Hist. 
Exact Sci.} {\bf 53} (1999) 529-575; arXiv:physics/9811050.

\noindent (Som 05) Klaus P. {\it Sommer} ``Nicht das Deutschland von
Hindenburg und Ludendorff'', sondern das von Hilbert und Einstein. Der
Fund von Briefen von Einstein, Planck, Born, Debye, Nernst,
Sommerfeld, Ehrenfest, Weyl, Courant und Althoff an David Hilbert auf
einem G\"ottinger Dachboden, {\it Berichte f\"ur
  Wissenschaftsgeschichte} (to be published).  

\noindent (Viz 01) V.P. {\it Vizgin}, On the discovery of the
gravitational field equations  by Einstein and Hilbert: new material,
{\it Uspekhi Fizicheskikh Nauk} {\bf 171} (2001) 1347-1363 (English
translation: {\it Physics-Uspekhi} {\bf 44} (2001) 1283-1298).

\noindent (Viz 94) V.P. {\it Vizgin}, {\it Unified Field Theories in
 the First Third of the 20th Century} (Birkh\"auser, Basel et al. 1994).

\noindent (Win 04) F. {\it Winterberg}, On ``Belated decision in the 
Hilbert-Einstein priority dispute'' published by L. Corry, J. Renn,
and J. Stachel, {\it Z. Naturforsch.} {\bf 59a} (2004) 715-719.  

\noindent (Wuensch 2005) Daniela {\it Wuensch}, {\it ``zwei wirkliche
  Kerle''} Neues
zur Entdeckung der Gravitationsgleichungen der Allgemeinen
Relativit\"atstheorie durch Albert Einstein und David Hilbert
(Termessos Verlag, G\"ottingen 2005).   

\noindent (Yan 02) Benjamin H. {\it Yandell}, {\it The Honors Class:
  Hilbert's Problems and Their Solvers} (A.K. Peters, Natik, MA 2002).

\vfill\eject
\bye

%% file: ictp.tex
%
\magnification=1200
\vsize=24truecm
 \nopagenumbers
 \footline={ \ifnum\pageno = 1 \primapag\else\altrepag\fi}
      \def\primapag{\hss\ \hss}
       \def\altrepag{\hss\folio\hss}

\def\TITLE#1{{\centerline{\bf #1}}
}

\def\AUTHOR#1{{\centerline {#1}}\smallskip}

\def\ABSTRACT{{\centerline{ABSTRACT}}\bigskip}

\parindent=40pt

\def\FOOTNOTE#1#2{\rm\parindent=0pt\footnote{#1}{#2}\parindent=40pt}
  \baselineskip=14pt
  \parskip=7pt plus 1pt

\def\today{\ifcase\month\or
             January\or February\or March \or April\or May\or June\or
             July\or August\or September\or October\or November\or December\fi
             \ \number\year}

\global\newcount\secno \global\secno=0
\global\newcount\meqno \global\meqno=1
\global\newcount\subsecno \global\subsecno=0
\def\SECTION#1{\global\advance\secno by1\global\meqno=1\global\subsecno=0
\bigbreak   \bigskip    
\noindent \hangindent 40pt {\bf\hbox to 40pt{\the\secno.\hfil}
           #1}\par\nobreak
                 \medskip\nobreak\message{#1 , }}
\def\SUBSECTION#1{\global\advance\subsecno by1\medbreak
      \noindent\hangindent 40pt
     {\bf\hbox to 40pt{\the\secno.\the\subsecno\hfil}#1}\par\nobreak
                 \medskip\nobreak\message{#1 ,  }}




\global\newcount\ftno \global\ftno=1
\def\FOOT#1{\footnote{$^{\the\ftno}$}{#1}\ %
\global\advance\ftno by1}

\global\newcount\figno \global\figno=1
\newwrite\ffile
\def\fig#1#2{\the\figno\nfig#1{#2}}
\def\nfig#1#2{\xdef#1{\the\figno}%
\ifnum\figno=1\immediate\openout\ffile=figs.tmp\fi%
\immediate\write\ffile {\noexpand \item{Fig. \noexpand#1 :\ }\noexpand#2}%
\global\advance\figno by1}
\def\semi{;\hfil\noexpand\break}

\global\newcount\tableno \global\tableno=1
\newwrite\ffile
\def\table#1#2{\the\tableno\ntable#1{#2}}
\def\ntable#1#2{\xdef#1{\the\tableno}%
\ifnum\tableno=1\immediate\openout\ffile=table.tmp\fi%
\immediate\write\ffile {\noexpand \item{Table. \noexpand#1 :\ }\noexpand#2}%
\global\advance\tableno by1}

\global\newcount\refno \global\refno=1
\newwrite\rfile
\def\ref#1#2{$^{[\the\refno]}$\nref#1{#2}}
\def\nref#1#2{\xdef#1{$^{[\the\refno]}$}%
\ifnum\refno=1\immediate\openout\rfile=refs.tmp\fi%
\immediate\write\rfile{\noexpand\item{\noexpand#1\ }\noexpand#2.}%
\global\advance\refno by1}




\def\ACK{\vskip 4truecm\centerline{\bf Acknowledgments}\par\nobreak
      \bigskip\nobreak}

\def\mc{\,\raise -2.truept\hbox{\rlap{\hbox{$\sim$}}\raise5.truept
\hbox{$<$}\ }}
\def\Mc{\,\raise -2.truept\hbox{\rlap{\hbox{$\sim$}}\raise5.truept
\hbox{$>$}\ }}
%
%

%
%

%
%

%
%
\def\square{\sqcap\kern-6pt\lower2.4pt\hbox{--}\ }
\def\sq{\sqcap\kern-8pt\lower2.4pt\hbox{--}\ }
%
%
\def\bbone{{\rm 1}\kern-4pt{\rm 1}}

\def\bbc{{\rm C}\kern-5pt\hbox{\vrule height6.5pt width0.8pt}\ \, }

\def\bbg{{\rm G}\kern-5pt\hbox{\vrule height6pt width 0.8pt}\ \, }

\def\bbo{{\rm O}\kern-4.8pt\hbox{\vrule height6.5pt width0.8pt}\  }

\def\bbq{{\rm Q}\kern-5pt\hbox{\vrule height6pt width 0.7pt}\  \, }

\def\bbs{{\rm S}\kern-3.5pt\hbox{\vrule height6.5pt width 0.7pt}\ }

\def\highchi{\raise 2pt\hbox{$\chi$}}
\def\subbbc{{\rm C}\kern-3.5pt\hbox{\vrule height4.5pt width0.4pt}\, }
%

%

%

%
%
%
\def\frac#1/#2{\leavevmode\kern.1em
\raise.5ex\hbox{\the\scriptfont0 #1}
\kern-.1em/\kern-.15em\lower.25ex\hbox{\the\scriptfont0 #2}}
%


%
\def\barh{h\kern-5pt\raise3pt\hbox{-}\ }
\def\ssbarh{h\kern-4.5pt\raise3pt\hbox{-}\,}
\def\ovssbarh{h\kern-3.5pt\hbox{-}\,}


\def\longrightharpoonup{-\kern-3pt\hbox{$\rightharpoonup$}\ }

\def\centerpar{
\let\endgraf=\par \edef\restorehsize{\hsize=14truecm}
\def\par{\endgraf \restorehsize \let\par=\endgraf}
\advance\hsize by-\parindent
\item{}}